\documentclass[]{raa}            
\usepackage{graphicx,times}
\usepackage{amssymb}
\usepackage{amsmath}
\usepackage{natbib}
\begin{document}
   \title{The rate of period change in DAV stars
}

\volnopage{ {\bf 0000} Vol.\ {\bf 0} No. {\bf XX}, 000--000}
\setcounter{page}{1}

\author{Y. H. Chen\inst{1,2,3} \ C. Y. Ding\inst{1,2,3} \ W. W. Na\inst{3,4} \ H. Shu\inst{1,2,3}}

\institute{\inst{1} Institute of Astrophysics, Chuxiong Normal University, Chuxiong 675000, China; {yhc1987@cxtc.edu.cn\\
           \inst{2} School of Physics and Electronical Science, Chuxiong Normal University, Chuxiong 675000,China\\
           \inst{3} Key Laboratory for the Structure and Evolution of Celestial Objects, Chinese Academy of Sciences, P.O. Box 110, Kunming 650011, China\\
           \inst{4} School of Physics and Electronic Engineering, Yuxi Normal University, Yuxi 653100,China}
\\
\vs \no
{\small Received [0000] [July] [day]; accepted [0000] [month] [day] }}

\abstract{Grids of DAV star models are evolved by \texttt{WDEC}, taking the element diffusion effect into account. The grid parameters are hydrogen mass log($M_{H}/M_{*}$), helium mass log($M_{He}/M_{*}$), stellar mass $M_{\rm *}$, and effective temperature $T_{\rm eff}$ for DAV stars. The core compositions are from white dwarf models evolved by \texttt{MESA}. Therefore, those DAV star models evolved by \texttt{WDEC} have historically viable core compositions. Based on those DAV star models, we studied the rate of period change ($\dot{P}(k)$) for different values of H, He, $M_{\rm *}$, and $T_{\rm eff}$. The results are consistent with previous work. Two DAV stars G117-B15A and R548 have been observed around forty years. The rates of period change of two large-amplitude modes were obtained through O-C method. We did asteroseismological study on the two DAV stars and then obtained a best-fitting model for each star. Based on the two best-fitting models, the mode identifications ($l$, $k$) of the observed modes for G117-B15A and R548 are consistent with previous work. Both the observed modes and the observed $\dot{P}$s can be fitted by calculated ones. The results indicate that our method of evolving DAV star models is feasible.
 }

\authorrunning{Y. H. Chen, C. Y. Ding, W. W. Na, H. Shu}            
\titlerunning{The rate of period change in DAV stars}  
\maketitle

\section{Introduction}

White dwarfs are the last observable stage of evolution for most of low and medium mass stars. They comprise around 98\% of the end state of all stars (Winget \& Kepler 2008). Thermonuclear reactions have basically stopped inside the white dwarf. The central core temperature gradually decreases with the residual heat continually radiating from the surface. The white dwarfs are darker and darker until we can not observe them. Those field white dwarfs with rich helium atmosphere are characterized as DO (strong He II lines) type or DB (strong He I lines) type white dwarfs, while those with rich hydrogen atmosphere are characterized as DA (only Balmer lines) type white dwarfs (McCook \& Sion 1999). Around 80\% (Bischoff-Kim \& Metcalfe 2011) of the population of field white dwarfs are DA type white dwarfs which have layered structure due to element diffusion.

In the Hertzsprung-Russel diagram, there are DOV (around 170000\,K to 75000\,K), DBV (around 29000\,K to 22000\,K), and DAV (around 12270\,K to 10800\,K) instability strips (Winget \& Kepler 2008). There are at least 148 DAV stars observed (Castanheira et al. 2010). The detailed DAV instability strip is closely related to the stellar gravity acceleration (Gianninas, Bergeron \& Fontaine 2005, Gianninas, Bergeron \& Ruiz 2011). The instability strip is significantly wide for DAV stars with large gravity accelerations. When DA type white dwarfs pass through the DAV instability trip, they will pulsate and change to be DAV stars. The DAV stars cool down slowly and therefore the cooling rate is very small. Cooling and contraction produced by evolution can affect the pulsation periods. Therefore, it is very important to study the changing rate of pulsation periods, which can reflect the evolution of a star.

The cooling of a white dwarf is a very long process. The rate of period change is very small and not easy to detect. Using the Taylor's formula in mathematics, astronomers have found a way to measure a minute rate of change, named O-C. The O-C method means the observation minus the calculation, which treats the time of maximal light as a function of its cycle number. The O-C method has been used to calculate the rate of period change in pre-white dwarf star PG 0122+200 (Fu et al. 2002), PG 1159-035 (Costa \& Kepler 2008), DBV star EC 20058-5234 (Dalessio et al. 2013), and DAV star G117-B15A (Kepler 2000, Kepler et al. 2005a, Kepler 2012), R548 (Mukadam et al. 2013). The rate of period change can be used to measure the evolutionary rate of white dwarf stars (Kepler et al. 2005b).

Bradley \& Winget (1991) discussed in detail the rate of period change in DBV and DAV stars with pure carbon cores. Studying the rate of period change in DAV stars with carbon/oxygen cores, Bradley, Winget \& Wood (1992) reported that the rates of period change for trapped modes were about half of that values for nontrapped modes. Bradley (1996) reported that the rate of period change was very sensitive to changes in the stellar mass and core compositions. The asteroseismological study including rate of period change were studied by Bradley (1998) for G117-B15A, R548 and by Bradley (2001) for L19-2, GD165. Bischoff-Kim, Montgomery \& Winget (2008a) did asteroseismological study on DAV star G117-B15A and R548. They (2008b) calculated the rate of period change for the best-fitting model of G117-B15A and tried to limit the axion mass. Giammichele et al. (2016) did asteroseismological study on R548 based on the optimization package LUCY (genetic evolLUtion Code for asteroseismologY). Based on new atmospheric modeling for DA white dwarfs, Giammichele et al. (2015) provided updated estimates of the time-averaged atmospheric properties of R548. The parameters of their best asteroseismological model are consistent with that of their best spectroscopical model, as shown in Table 8 of Giammichele et al. (2016). The calculated rate of period change for the 213\,s mode is consistent with the observed value obtained through O-C.

For model calculations, ($k,l,m$) are used to characterize an eigen-mode. The three indices are the radial order, the spherical harmonic degree, and the azimuthal number, respectively. For DAV stars, the eigen-modes are non-radial $g$-mode pulsations. The rate of period change for pulsation periods with same $l$ value (usually $l$ = 1 or 2) and same $m$ value ($m$ = 0) can be calculated by
\begin{equation}
\dot{P}(k)=\frac{P_{2}(k)-P_{1}(k)}{Age_{2}-Age_{1}}.
\end{equation}
\noindent In Eq.\,(1), $P_{2}(k)$ is the pulsation period of a DAV star with stellar age to be $Age_{2}$ and $P_{1}(k)$ is the pulsation period of the DAV star with stellar age to be $Age_{1}$. Precisely, $P_{2}(k)$ and $P_{1}(k)$ are the period of the same mode (same $k$) at two different epochs. $\dot{P}(k)$ can represent the corresponding rate of period ($k$) change.

\texttt{MESA} is a stellar evolution code reported by Paxton et al. (2011, 2013). A main-sequence ($MS$) star can be evolved to be a white dwarf ($WD$) by \texttt{MESA}. The core compositions of the white dwarf are results of thermal nuclear burning. However, the DAV stars evolved by \texttt{MESA} usually have rich He/H envelopes. It is not convenient for \texttt{MESA} to evolve grids of DAV star models. \texttt{WDEC} is a quasi-static code to calculate the cooling process of white dwarf stars (Montgomery et al. 1999). Before the evolution, \texttt{WDEC} is convenient to enter the hydrogen layer mass fraction (log($M_{H}/M_{*}$)), the helium layer mass fraction (log($M_{He}/M_{*}$)), the total stellar mass ($M_{\rm *}$), and the effective temperature ($T_{eff}$) for the output white dwarfs. Namely, \texttt{WDEC} is suitable for evolving grids of white dwarf models. However, the core compositions are usually artificial, such as full carbon (C), full oxygen (O), or homogeneous C/O core compositions. Adding the core compositions of white dwarfs evolved by \texttt{MESA} into \texttt{WDEC}, grids of white dwarf models can be evolved by \texttt{WDEC}. Those white dwarf models have historically viable core composition profiles. The element diffusion scheme of Thoul, Bahcall \& Loeb (1994) is added into \texttt{WDEC} by Su et al. (2014). The method has been used many times recently to evolve DAV stars (Chen \& Li 2014a, Chen \& Li 2014b, Su et al. 2014, Chen 2016a) and DBV stars (Chen 2016b). In this paper, we studied the rate of period change in DAV stars based on the DAV star models evolved in the method. Based on observations of about forty years on G117-B15A and R548, the rates of period change for two large-amplitude modes are obtained through O-C method (Kepler 2012, Mukadam et al. 2013). We try to check the DAV star models by doing asteroseismological study on the two DAV stars and comparing the calculated rates of period change to the observed values.

In Sect. 2, we briefly introduce the input physics and model calculations and discuss the $\dot{P}(k)$ values with various differences of stellar ages. We studied the rate of period change in DAV stars in Sect. 3, taking different values of H, He, $M_{\rm *}$, and $T_{eff}$ into account. In Sect. 4, we did the asteroseismological study on DAV stars G117-B15A, R548, and then calculate the rates of period change for the two best-fitting models. The theoretical values of $\dot{P}(k)$ can be compared with the observed values obtained through the O-C method. At last, we give a discussion and conclusions in Sect. 5.

\section{Input physics and model calculations}

The 6208 version of \texttt{MESA} is downloaded and installed. In the module of 'make\_co\_wd', the initial $MS$ stellar masses are entered, as shown in the first column of Table 1. The other input is the default values. The initial metal abundance is 0.02 and the mixing length parameter is 2.0. When the logarithm of stellar luminosity divided by solar luminosity is less than -2.0, those $MS$ stars have evolved to be $WD$ stars. The corresponding $WD$ masses are shown in the second column of Table 1. Those $WD$s usually have rich He/H envelopes. The core structure parameters, including mass, radius, luminosity, pressure, temperature, entropy and C profile, are took out and added into \texttt{WDEC}. \texttt{WDEC} was first developed by Martin Schwarzschild and subsequently modified by Kutter \& Savedoff (1969), Lamb \& van Horn (1975), and Wood (1990). With opacities of Itoh et al. (1983, 1984), the equation of state is from Lamb (1974) and Saumon, Chabrier \& van Horn (1995). The standard mixing length theory is used and the mixing length parameter is adopted as ML2/$\alpha$=0.6 for DAV stars. The core boundary of $WD$s evolved by \texttt{MESA} is defined around the peak of the C abundance. The O abundance equals one unit minus the C abundance. The corresponding core masses are shown in the third column of Table 1. With historically viable core compositions, grids of $WD$s can be evolved by \texttt{WDEC}. The composition profiles are results of diffusion, not the previous approximations of diffusion equilibrium H/He and He/C transitions (Su et al. 2014). The corresponding $WD$ masses evolved by \texttt{WDEC} are shown in the fourth column of Table 1. The WD masses in the fourth column are around the values in the third column. It is an approximation in order to evolve more dense WD masses by \texttt{WDEC}. With a modified pulsation code of Li (1992a,b), full equations of linear and adiabatic oscillation are solved and the eigen-frequencies can be found one by one through scanning. Then, Eq.\,(1) can be used to calculate the rate of period change.

In order to calculate $\dot{P}(k)$, two models with different ages should be selected. We choose typical DAV stars with log($M_{H}/M_{*}$) = -4.0, log($M_{He}/M_{*}$) = -2.0, $M_{\rm *}$ = 0.600\,$M_{\odot}$, and $T_{eff}$ around 12000\,K. Four groups of DAV stars with different stellar ages are chosen, as shown in Fig. 1. The abscissa is pulsation period $P(k)$ which is from 0\,s to 1500\,s. The ordinate is $\dot{P}(k)$. The pluses, boxes, forks, and circles respectively correspond to DAV star models of $T_{eff}$ = 12050\,K and 11950\,K, 12100\,K and 11900\,K, 12200\,K and 11800\,K, 12300\,K and 11700\,K. Then, the values of $\dot{P}(k)$ can be calculated according to Eq.\,(1). The radial order $k$ is from 1 to 31 in the period range for those models. In Fig. 1, we can see that the values of $\dot{P}(k)$ are basically equaled for $\Delta$$T_{\rm eff}$ = 100\,K, 200\,K, 400\,K, and 600\,K. We choose models of $\Delta$$T_{\rm eff}$ = 400\,K to calculate the theoretical values of $\dot{P}(k)$ in this paper.

\begin{table}
\caption{Masses of $MS$ stars, $WD$ stars, corresponding cores evolved by \texttt{MESA}, and masses of corresponding $WD$ stars evolved by \texttt{WDEC}.}
\begin{center}
\begin{tabular}{lllll}
\hline
$MS$         &$WD(\texttt{MESA})$   &$M_{core}(\texttt{MESA})$ &$WD(\texttt{WDEC})$     \\
\hline
[$M_{\odot}$]&[$M_{\odot}$]         &[$M_{\odot}$]             &[$M_{\odot}$]           \\
\hline
3.00         &0.579                 &0.550                     &0.550-0.560             \\
3.20         &0.599                 &0.575                     &0.565-0.585             \\
3.40         &0.627                 &0.600                     &0.590-0.610             \\
3.50         &0.652                 &0.625                     &0.615-0.635             \\
3.60         &0.675                 &0.650                     &0.640-0.660             \\
3.70         &0.694                 &0.675                     &0.665-0.685             \\
3.80         &0.714                 &0.770                     &0.690-0.710             \\
3.92         &0.738                 &0.725                     &0.715-0.735             \\
4.00         &0.763                 &0.750                     &0.740-0.750             \\
\hline
\end{tabular}
\end{center}
\end{table}

\begin{figure}
\begin{center}
\includegraphics[width=9.0cm,angle=0]{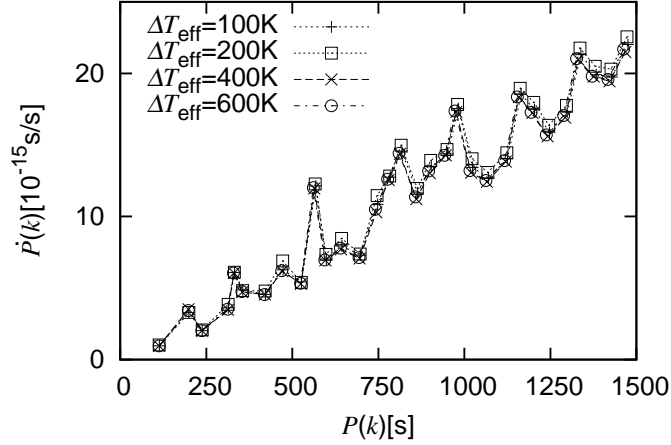}
\end{center}
\caption{The rate of period change for models with different effective temperatures. The rate of period change is calculated by the difference of two eigen-periods (same $k$ values) divided by the difference of corresponding model ages. The parameters of the models are log($M_{\rm H}/M_{\rm *}$) = -4.0, log($M_{\rm He}/M_{\rm *}$) = -2.0, $M_{\rm *}$ = 0.600\,$M_{\odot}$, and $T_{\rm eff}$ around 12000\,K. For example, the effective temperature is 12050\,K and 11950\,K for pluses and then $\Delta$$T_{\rm eff}$ is 100\,K.}
\end{figure}

Some groups of DAV star models are evolved in order to study the rate of period change in DAV stars. For example, in order to study the effect of different values of H, we evolve DAV star models of log($M_{H}/M_{*}$) from -10.0 to -4.0 with steps of 1.0, log($M_{He}/M_{*}$) = -2.0, $M_{\rm *}$ = 0.600\,$M_{\odot}$, and $T_{eff}$ = 12200\,K and 11800\,K. In order to do the asteroseismological study on DAV star G117-B15A and R548, grids of DAV star models are evolved. The grid parameters are log($M_{H}/M_{*}$) from -10.0 to -4.0 with steps of 1.0, log($M_{He}/M_{*}$) from -4.0 to -2.0 with steps of 0.5, $M_{\rm *}$ from 0.550\,$M_{\odot}$ to 0.750\,$M_{\odot}$ with steps of 0.010\,$M_{\odot}$, and $T_{eff}$ from 12800\,K to 10800\,K with steps of 100\,K. After pre-selecting a pre-best-fitting model, the grids of DAV star models will be made dense around the pre-best-fitting parameters. Then, a best-fitting model will be selected. At last, the calculated $\dot{P}(k)$ of the best-fitting model can be compared with corresponding observed $\dot{P}$ through the O-C method.

\section{The rate of period change in DAV stars}

With the evolved DAV star models, we studied the rate of period change in DAV stars. The Eq.\,(1) is used on two DAV star models with differences of $T_{eff}$ to be 400\,K. The pulsation code starts to calculate eigen-modes from $k$ = 1. The lower $k$ $g$-modes of each model will be calculated. We studied the effect of different values of H, He, kinetic energy distribution, and different values of $M_{\rm *}$, $T_{eff}$ on the rate of period change. It may be helpful for qualitative analysis on the rate of period change to some observed star.

 \subsection{The effect of different values of H atmosphere mass and He layer mass}

\begin{figure}
\begin{center}
\includegraphics[width=9.0cm,angle=0]{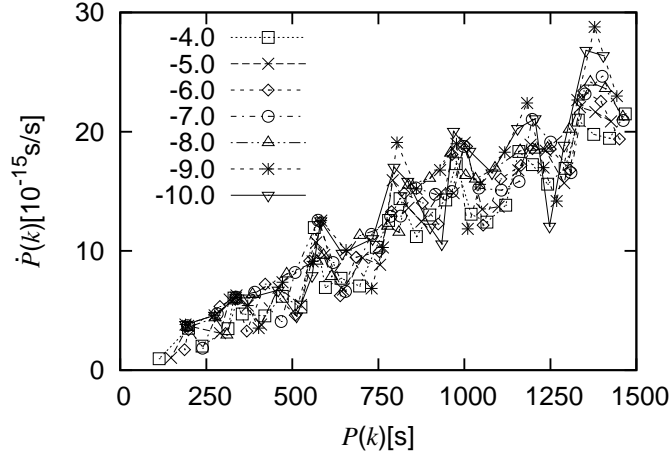}
\end{center}
\caption{The effect of different H atmosphere masses on the rate of period change. The parameters of the models are log($M_{\rm H}/M_{\rm *}$) from -10.0 to -4.0 with steps of 1.0, log($M_{\rm He}/M_{\rm *}$) = -2.0, $M_{\rm *}$ = 0.600\,$M_{\odot}$, and $T_{\rm eff}$ = 12200\,K and 11800\,K.}
\end{figure}

\begin{figure}
\begin{center}
\includegraphics[width=9.0cm,angle=0]{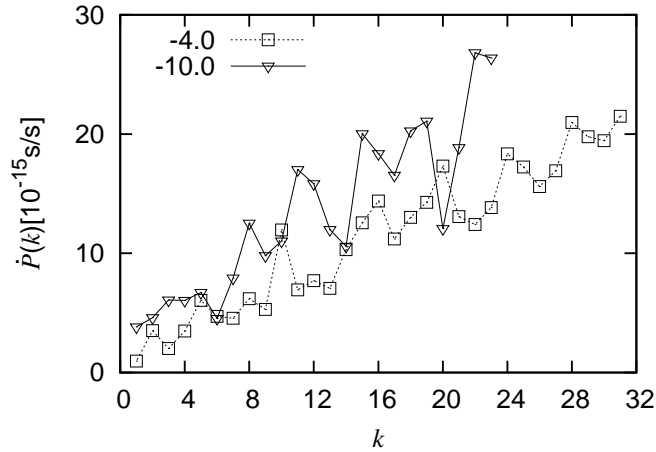}
\end{center}
\caption{The effect of different H atmosphere masses on the rate of period change. The abscissa is the radial order $k$. The squares and triangles are values for log($M_{\rm H}/M_{\rm *}$).}
\end{figure}

\begin{figure}
\begin{center}
\includegraphics[width=9.0cm,angle=0]{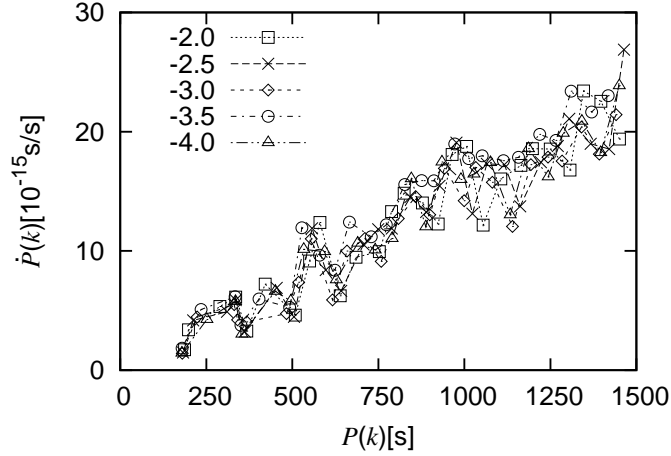}
\end{center}
\caption{The effect of different He layer masses on the rate of period change. The parameters of the models are log($M_{\rm H}/M_{\rm *}$) = -6.0, log($M_{\rm He}/M_{\rm *}$) from -4.0 to -2.0 with steps of 0.5, $M_{\rm *}$ = 0.600\,$M_{\odot}$, and $T_{\rm eff}$ = 12200\,K and 11800\,K.}
\end{figure}

First of all, we studied the effect of different values of H atmosphere mass and He layer mass on the rate of period change in DAV stars. In Fig. 2, we show a diagram of $\dot{P}(k)$ to $P(k)$ with different values of H atmosphere mass. With $T_{\rm eff}$ = 12200\,K and 11800\,K respectively, log($M_{\rm He}/M_{\rm *}$) is fixed to be -2.0, $M_{\rm *}$ is fixed to be 0.600\,$M_{\odot}$, and log($M_{\rm H}/M_{\rm *}$) is from -10.0 to -4.0 with steps of 1.0. In the period range from 0\,s to 1500\,s, the radial order $k$ is from 1 to 31 for the model of log($M_{\rm H}/M_{\rm *}$) = -4.0 and from 1 to 23 for the model of log($M_{\rm H}/M_{\rm *}$) = -10.0. The pulsation period is from 113.16\,s ($k$ = 1) to 1467.93\,s ($k$ = 31) for the model of log($M_{\rm H}/M_{\rm *}$) = -4 and $T_{\rm eff}$ = 12200\,K. For models of log($M_{\rm H}/M_{\rm *}$) = -4, $T_{\rm eff}$ = 12200\,K and 11800\,K, the differences of corresponding pulsation periods are from 0.92\,s to 20.53\,s and the difference of ages is 0.30 $\times$ $10^{8}$ years. The rate of period change is from 0.97 to 21.70 in a unit of $10^{-15}$s/s, marked as boxes in Fig. 2.

The values of $\dot{P}(k)$ are all positive. This is because the cooling process dominates for cool white dwarfs, which will increase the degree of degeneracy and increase the pulsation period (Winget, Hansen \& van Horn 1983, Kepler et al. 2000, Winget \& Kepler 2008). Overall, the rates of period change for different H atmosphere mass models are basically overlapped with each other. We notice that the rate of period change is distributed in a strip from the lower left corner to the upper right corner. Basically, the longer the pulsation period, the larger the rate of period change of a mode. This is because that the asymptotic period spacing is increasing with the white dwarf cooling down. The asymptotic period spacing (Tassoul 1980) can be calculated by
\begin{equation}
\bar{\triangle \texttt{$P$}(l)}=\frac{2\pi^{2}}{\sqrt{l(l+1)}{\int_{0}}^{R}\frac{|N|}{r}dr}.
\end{equation}
\noindent In Eq.\,(2), $R$ is stellar radius and $N$ is Brunt-V\"ais\"al\"a frequency. Therefore, we basically have equations of
\begin{equation}
\begin{split}
    &  \quad P(k)=P(k_{0})+(k-k_{0})*\triangle \bar{\texttt{$P$}},\\
and &  \quad \dot{P}(k)=\dot{P}(k_{0})+(k-k_{0})*\triangle \dot{\bar{\texttt{$P$}}}.
\end{split}
\end{equation}
\noindent We assume $k_{0}$ is so high that the value of $P(k_{0})$ satisfies the asymptotic period spacing law. Therefore, periods with high $k$ values will increase with both the cooling process of a DAV star and the increasing process of the asymptotic period spacing. The long-period modes should have relatively large rate of period change. In addition, some values of $\dot{P}(k)$ are floating around the strip. For example, the inverted triangle is relatively large around 1000\,s and relatively small around 1250\,s. The effect is associated with mode trapping effect, which will be discussed in the next subsection.

The results are consistent with that of Bradley \& Winget (1991). They found that the rate of period change for the model of log($M_{\rm H}/M_{\rm *}$) = -4.0 was about half of that for the model of log($M_{\rm H}/M_{\rm *}$) = -10.0. The conclusion is based on the abscissa to be radial order $k$. However, the abscissa in Fig. 2 is the pulsation period. In Fig. 3, we show the rate of period change to the radial order $k$. We can see that the values of $\dot{P}(k)$ for the model of log($M_{\rm H}/M_{\rm *}$) = -4.0 are obviously smaller than that for the model of log($M_{\rm H}/M_{\rm *}$) = -10.0. The frequency of an eigen-mode can be obtained from observations but the radial order $k$ can not be obtained. Therefore, diagrams of $\dot{P}(k)$ to $P(k)$ are displayed in Fig 2.

In Fig. 4, we show a diagram of $\dot{P}(k)$ to $P(k)$ with different values of He layer mass. For the models, log($M_{\rm H}/M_{\rm *}$) is -6.0, log($M_{\rm He}/M_{\rm *}$) is from -4.0 to -2.0 with steps of 0.5, $M_{\rm *}$ is fixed to be 0.600\,$M_{\odot}$, and $T_{\rm eff}$ equals 12200\,K and 11800\,K respectively. The radial order $k$ is from 1 to 27 for the model of log($M_{\rm He}/M_{\rm *}$) = -2.0 and from 1 to 25 for the model of log($M_{\rm He}/M_{\rm *}$) = -4.0. The values of $\dot{P}(k)$ are not sensitive to different He layer masses, which are consistent with that of Bradley \& Winget (1991). They are distributed in a strip from lower left corner to the upper right corner in Fig. 4. There are also dispersion of some $\dot{P}(k)$ values around the strip.

 \subsection{The effect of kinetic energy distribution}

\begin{figure}
\begin{center}
\includegraphics[width=9.0cm,angle=0]{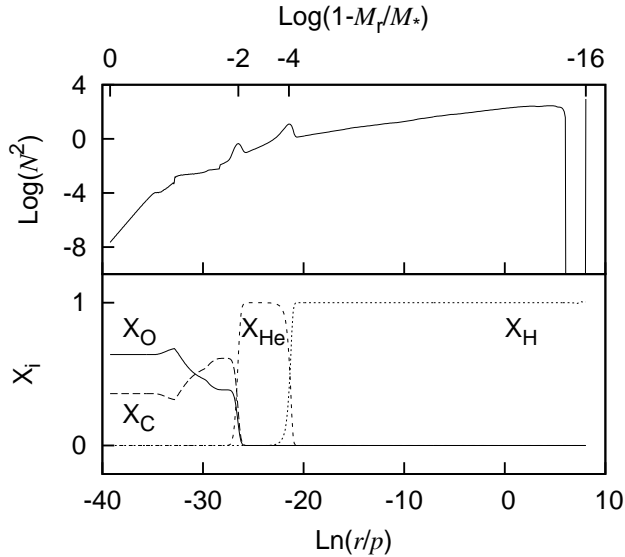}
\end{center}
\caption{The core composition profiles and the Brunt-V\"ais\"al\"a frequency for the model of log($M_{\rm H}/M_{\rm *}$) = -4.0, log($M_{\rm He}/M_{\rm *}$) = -2.0, $M_{\rm *}$ = 0.600\,$M_{\odot}$, and $T_{\rm eff}$ = 11800\,K.}
\end{figure}

\begin{figure}
\begin{center}
\includegraphics[width=9.0cm,angle=0]{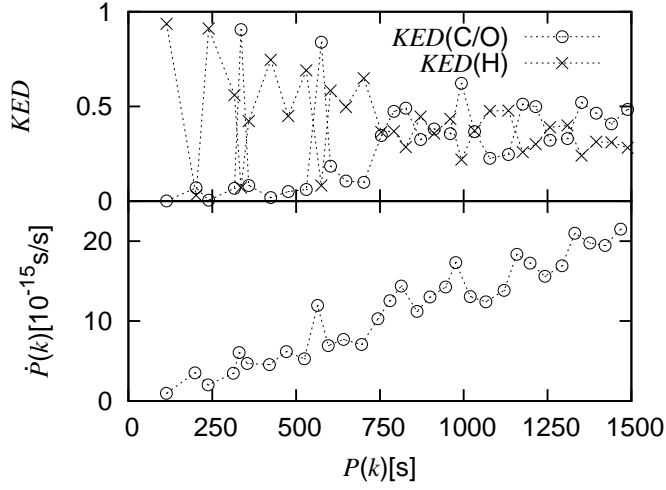}
\end{center}
\caption{Diagram of the rate of period change to pulsation periods and corresponding kinetic energy distributions. The model in the upper panel is the same with which in Fig. 5. The value $k$ is from 1 to 31. In the lower panel, two models of $T_{\rm eff}$ = 12200\,K and 11800\,K are used to calculate values of $\dot{P}(k)$.}
\end{figure}

\begin{figure}
\begin{center}
\includegraphics[width=9.0cm,angle=0]{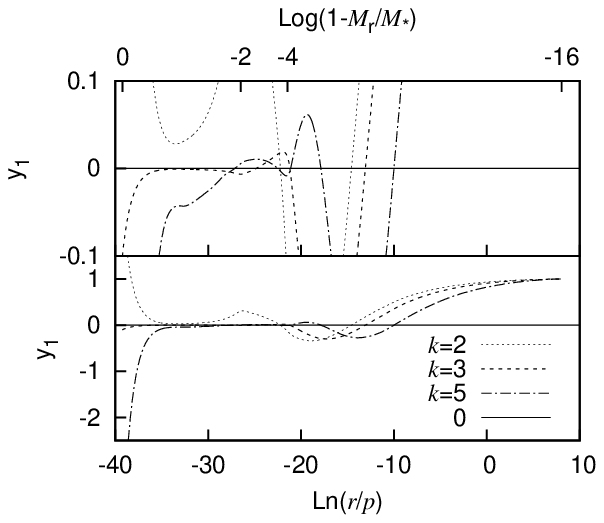}
\end{center}
\caption{Diagram of $y_{1}$ to Ln(r/p) for the modes of $k$ = 2, 3, and 5. The model parameters are the same with which in Fig. 5. The upper panel is a longitudinal enlargement of the lower one. The top abscissa is Log(1-$M_{\rm r}/M_{\rm *}$), which shows the regions of C/O core, He layer, and H atmosphere clear.}
\end{figure}

Both different H models and different He models are found to have $\dot{P}(k)$s floating around a strip. In this subsection, we try to discuss the rate of period change for each mode by analyzing its kinetic energy distribution ($KED$). In Fig. 5, we show the core composition profiles and the Brunt-V\"ais\"al\"a frequency of a DAV star model. The model parameters are log($M_{\rm H}/M_{\rm *}$) = -4.0, log($M_{\rm He}/M_{\rm *}$) = -2.0, $M_{\rm *}$ = 0.600\,$M_{\odot}$, and $T_{\rm eff}$ = 11800\,K. The C profile is from a 0.627\,$M_{\odot}$ white dwarf (0.600 \,$M_{\odot}$ core) evolved by \texttt{MESA} from a 3.40\,$M_{\odot}$ $MS$ star, as shown in Table 1. The white dwarfs evolved by \texttt{MESA} usually have rich He/H envelops. In the core, the O profile equals 1 minus the C profile. The H/He composition gradient zone and He/C/O composition gradient zone are results of element diffusion. The composition transition zone can make a spike in the Brunt-V\"ais\"al\"a frequency. The spikes in the upper panel in Fig. 5 will make mode trapping effect (Winget, van Horn \& Hansen 1981, Brassard et al. 1992). In addition, there is a thin convection zone on the surface in the upper panel.

In Fig. 6, we show the rate of period change and corresponding kinetic energy distributions of each mode. The model parameters are the same with that in Fig. 5. In order to calculate the rate of period change, models of $T_{\rm eff}$ = 11800\,K and 12200\,K are evolved. The pulsation periods and corresponding kinetic energy distributions are calculated for the model of $T_{\rm eff}$ = 11800\,K. The radial order $k$ is from 1 to 31. The kinetic energy distributions can be calculated by,
\begin{equation}
\begin{split}
 & \quad KED(C/O) \quad / \quad KED(He) \quad / \quad KED(H)\\
=& \quad {4\pi\int_{0}^{R_{He/C/O}}[(|\xi_{r}|^2+l(l+1)|\xi_{h}|^2)]\rho_{0}r^2dr} \\
/& \quad {4\pi\int_{R_{He/C/O}}^{R_{H/He}}[(|\xi_{r}|^2+l(l+1)|\xi_{h}|^2)]\rho_{0}r^2dr}\\
/& \quad {4\pi\int_{R_{H/He}}^{R}[(|\xi_{r}|^2+l(l+1)|\xi_{h}|^2)]\rho_{0}r^2dr}.
\end{split}
\end{equation}
\noindent In Eq.\,(4), $\rho_{0}$ is the local density. $KED(C/O)$, $KED(He)$, and $KED(H)$ respectively represents the value of kinetic energy for a given mode distributed in C/O core, He layer, and H atmosphere. $R_{He/C/O}$, $R_{H/He}$, and $R$ is the location of He/C/O interface, H/He interface, and stellar radius, respectively. For the model in Fig. 5 and 6, $R_{He/C/O}$ is the location of log(1-$M_{\rm r}/M_{\rm *}$) = -2.0 and $R_{H/He}$ is the location of log(1-$M_{\rm r}/M_{\rm *}$) = -4.0. As shown, for instance, in Christensen-Dalsgaard (2008), $\xi_{r}(r)$ is the radial displacement and $\xi_{h}(r)$ is the horizontal displacement. In the upper panel in Fig. 6, total kinetic energy is set to be 1. $KED(C/O)$ is represented by open dots and $KED(H)$ is represented by forks. Therefore, $KED(He)$ equals 1 minus $KED(C/O)$ and $KED(H)$.

The radial order $k$ is from 1 to 31 for the model, which can be counted one by one in Fig. 6. For the mode of $k$ = 1, 3, 7, 9, and 13, $KED(H)$ is obviously larger than 50\%. It means that those modes are trapped or partly trapped in H atmosphere. They have minimal rates of period change in the lower panel. For the model of $k$ = 2, both $KED(C/O)$ and $KED(H)$ are very small. Therefore, $KED(He)$ is relatively large which means that the mode is trapped or partly trapped in the He layer. In the lower panel, $\dot{P}(k=2)$ is a maximal one. For the mode of $k$ = 5, 10, and 20, $KED(C/O)$ is obviously larger than 50\%. They are trapped or partly trapped in the C/O core and correspond to the maximal rates of period change in the lower panel. Namely, modes trapped or partly trapped in C/O core or He layer have maximal $\dot{P}(k)$s. Modes trapped or partly trapped in H atmosphere have minimal $\dot{P}(k)$s. This is reasonable. Because the cooling rate of the C/O core and He layer is faster than that of the H atmosphere. The cooling process dominates for DAV stars (Winget, Hansen \& van Horn 1983). Bradley, Winget \& Wood (1992) reported that the rates of period change for trapped modes (modes trapped or partly trapped in H atmosphere) are about half the values for nontrapped modes. Different values of H atmosphere mass and He layer mass will affect the rate of period change by mode trapping effect.

In Fig. 7, we show a diagram of a variable $y_{1}$ versus the stellar structure parameter ln(r/p). In the abscissa, $r$ is stellar local radius and $p$ is stellar local pressure. The variable $y_{1}$ is a dimensionless variable of the radial displacement $\xi_{r}$,
\begin{equation}
y_{1}=\frac{GM}{R^{3}}\frac{\xi_{r}}{g_{r}}.
\end{equation}
\noindent In Eq.\,(5), $G$ is the gravitational constant, $M$ is stellar mass, and $g_{r}$ is stellar local gravitation. The modes of $k$ = 2, 3, and 5 are shown in the lower panel in Fig. 7. The upper panel is the enlargement of the lower one. The values of $k$ can be counted out from the intersection points between $y_{1}$ and zero lines. The abscissa on the upper boundary is log(1-$M_{\rm r}/M_{\rm *}$). The modes of $k$ = 2, 3, and 5 have their largest $y_{1}$ amplitude in the He layer, H atmosphere, and C/O core, respectively. The top abscissa in Fig. 7 shows the regions clear.

 \subsection{The effect of different values of total stellar mass and effective temperature}

\begin{figure}
\begin{center}
\includegraphics[width=9.0cm,angle=0]{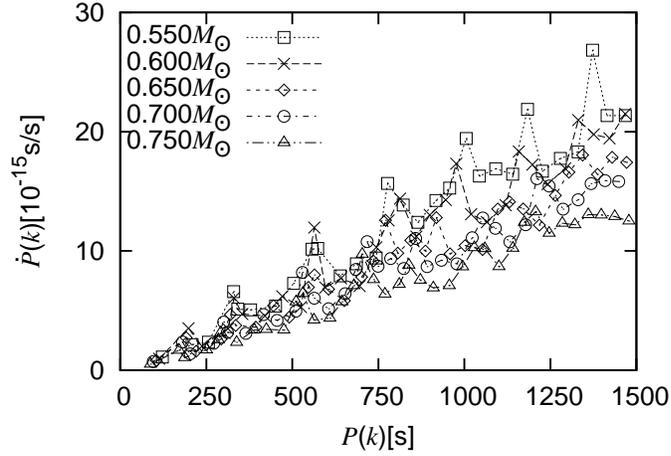}
\end{center}
\caption{The effect of different stellar masses on the rate of period change. The parameters of the models are log($M_{\rm H}/M_{\rm *}$) = -4.0, log($M_{\rm He}/M_{\rm *}$) = -2.0, $M_{\rm *}$ from 0.550\,$M_{\odot}$ to 0.750\,$M_{\odot}$ with steps of 0.050\,$M_{\odot}$, and $T_{\rm eff}$ = 12200\,K and 11800\,K.}
\end{figure}

\begin{figure}
\begin{center}
\includegraphics[width=9.0cm,angle=0]{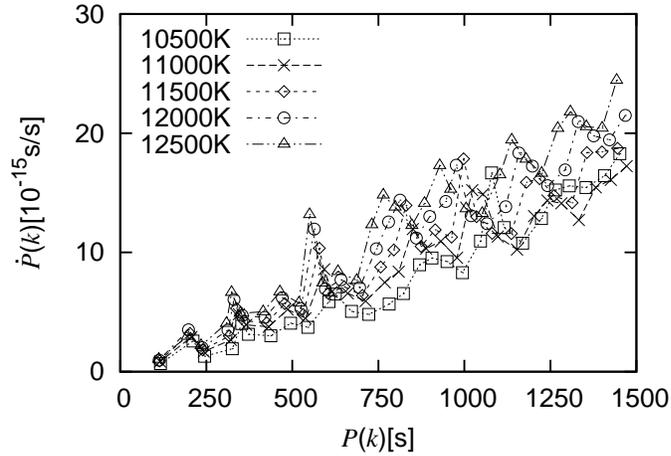}
\end{center}
\caption{The effect of different effective temperatures on the rate of period change. The parameters of the models are log($M_{\rm H}/M_{\rm *}$) = -4.0, log($M_{\rm He}/M_{\rm *}$) = -2.0, $M_{\rm *}$ = 0.600\,$M_{\odot}$, and $T_{\rm eff}$ from 12500\,K ($\pm$200\,K) to 10500\,K ($\pm$200\,K) with steps of 500\,K.}
\end{figure}

\begin{figure}
\begin{center}
\includegraphics[width=9.0cm,angle=0]{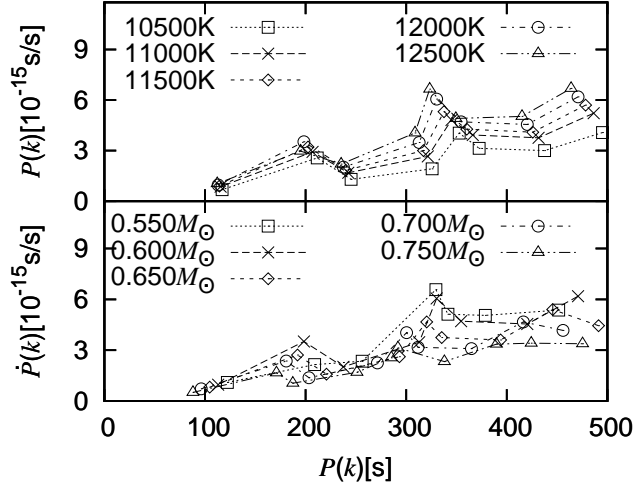}
\end{center}
\caption{The enlargement of Fig. 8 and Fig. 9 for short-period modes.}
\end{figure}

On the above discussion, we understand that the mode trapping effect will affect the rate of period change up and down. In this subsection, we will discuss the effect of different values of $M_{\rm *}$ and $T_{eff}$ on the rate of period change.

In Fig. 8, we show a diagram of $\dot{P}(k)$ to $P(k)$ with different $M_{\rm *}$ values. The model parameters are log($M_{\rm H}/M_{\rm *}$) = -4.0, log($M_{\rm He}/M_{\rm *}$) = -2.0, $M_{\rm *}$ from 0.550\,$M_{\odot}$ to 0.750\,$M_{\odot}$ with steps of 0.050\,$M_{\odot}$, and $T_{\rm eff}$ = 12200\,K and 11800\,K. We can see that the larger the total stellar mass of a DAV star, the smaller the rate of period change of its modes, especially for the long-period modes. On the one hand, a white dwarf with large stellar mass has a long process of cooling down and therefore a small cooling rate. A small cooling rate leads to a small rate of period change. On the other hand, a white dwarf with large stellar mass has a large gravitational acceleration and a small asymptotic period spacing according to Eq.\,(2). The radial order $k$ is from 1 to 29 for the model of $M_{\rm *}$ = 0.550\,$M_{\odot}$ and from 1 to 38 for the model of $M_{\rm *}$ = 0.750\,$M_{\odot}$ in the period range of 0-1500\,s. Therefore, the asymptotic period spacing is small for the model of $M_{\rm *}$ = 0.750\,$M_{\odot}$. Both the long process of cooling down and the small value of asymptotic period spacing lead to obviously small rate of period change for long-period modes of large stellar mass DAV stars.

In Fig. 9, we show a diagram of $\dot{P}(k)$ to $P(k)$ with different $T_{eff}$ values. The effective temperature is from 12500\,K to 10500\,K with steps of 500\,K. The rate of period change with $T_{eff}$ = 12500\,K is calculated from two models of $T_{eff}$ = 12700\,K and 12300\,K, namely $\Delta$$T_{\rm eff}$ = 400\,K. We can see that the higher the effective temperature, the larger the rate of period change. This is because that the hot DAV star cools faster. The radial order $k$ is from 1 to 29 for the model of $T_{eff}$ = 10500\,K and from 1 to 31 for the model of $T_{eff}$ = 12500\,K. Therefore, the change of asymptotic period spacing has small effect on the rate of period change for models of different $T_{\rm eff}$ values. Different cooling rates for different DAV star models dominate the values of $\dot{P}(k)$.

For the long-period modes, the effect of different values of $M_{\rm *}$ and $T_{eff}$ is relatively obvious. If we observe a large $\dot{P}(k)$ for a long-period mode, the star may have a small $M_{\rm *}$ or a high $T_{eff}$. The short-period modes in Fig. 8 and 9 are enlarged in Fig. 10. In the upper panel in Fig. 10, both the different $T_{eff}$ values and different kinetic energy distributions have an effect on the rate of period change. While, in the lower panel, different kinetic energy distributions dominate the distributions of $\dot{P}(k)$.

\section{The rate of period change in G117-B15A and R548}

Based on the discussions above, the values of $\dot{P}(k)$ are determined by $KED$ of $P(k)$, the stellar mass $M_{\rm *}$, and the effective temperature $T_{eff}$. In this section, we studied the rate of period change in DAV stars G117-B15A and R548. Grids of DAV star models are evolved. The pre-grid parameters are shown in the last paragraph of Sect. 2. The eigen-frequencies are scanned out among the grids of DAV star models and are used to fit the observed modes of G117-B15A and R548. A pre-best-fitting model is selected by,
\begin{equation}
\chi=\sqrt{\frac{\sum_{1}^{n_{obs}}(P_{\rm cal}-P_{\rm obs})^{2}}{n_{obs}}}.
\end{equation}
\noindent In Eq.\,(6), $P_{\rm cal}$ is the calculated modes, $P_{\rm obs}$ is the observed modes, and $n_{obs}$ is the number of observed modes. The model of minimal $\chi$ is selected as the pre-best-fitting one. Then, the grids of DAV star models will be made dense around the pre-best-fitting parameters. Then, a best-fitting model will be selected according to Eq.\,(6). We will calculate the rate of period change of modes from the best-fitting model and compare them with corresponding observed rate of period change obtained through O-C method.

 \subsection{Asteroseismological study on DAV star G117-B15A and R548}

\begin{table}
\caption{The observed modes from Table 5 of Romero et al. (2012) and calculated modes from corresponding best-fitting models. The value of $P_{\rm obs}$ minus $P_{\rm cal}$ and the parameter $\chi$ are also displayed. For R548, a new mode of 217.83\,s was identified by Giammichele et al. (2015).}
\begin{center}
\begin{tabular}{lllll}
\hline
Star         &$P_{\rm obs}$($l$)         &$P_{\rm cal}$($l$,$k$)   &$P_{\rm obs}$-$P_{\rm cal}$        &$\chi$ \\
\hline
             &[s]                        &[s]                      &[s]                                &[s]    \\
\hline
G117-B15A    &215.20(1)                  &213.86(1,1)              &$\,$ 1.34                          &1.32   \\
             &270.46(1)                  &270.92(1,2)              &$\,$-0.46                          &       \\
             &304.05(1)                  &305.85(1,3)              &$\,$-1.80                          &       \\
\hline
R548         &187.28(1or2)               &190.42(2,3)              &$\,$-3.14                          &2.12   \\
             &212.95(1)                  &214.41(1,1)              &$\,$-1.46                          &       \\
             &274.51(1)                  &272.48(1,2)              &$\,$ 2.03                          &       \\
             &318.07(1or2)               &317.97(1,3)              &$\,$ 0.10                          &       \\
             &333.64(1or2)               &331.13(1,4)              &$\,$ 2.51                          &       \\
\hline
R548         &217.83(2)                  &218.82(2,4)              &$\,$-0.99                          &1.97   \\
\hline
\end{tabular}
\end{center}
\end{table}

\begin{table}
\caption{The parameters of ten fitting models with smallest $\chi$s fitting G117-B15A and R548.}
\begin{center}
\begin{tabular}{lcccccccccccc}
\hline
                      &log($M_{H}/M_{*}$)  &log($M_{He}/M_{*}$)  &$T_{\rm eff}$       &$M_{\rm *}$  &log$g$ &$\chi$ \\
\hline
                      &                    &                     &[K]                 &[$M_{\odot}$]&       &[s]    \\
\hline
ID(G117-B15A)         &                    &                     &                    &             &       &       \\
\hline
$\,$ model1           &-8.0                &-3.0                 &11900               &0.660        &8.1912 &1.32   \\
$\,$ model2           &-8.0                &-3.0                 &11950               &0.660        &8.1911 &1.33   \\
$\,$ model3           &-8.0                &-3.0                 &11750               &0.665        &8.1940 &1.44   \\
$\,$ model4           &-8.0                &-3.0                 &11850               &0.660        &8.1913 &1.48   \\
$\,$ model5           &-8.0                &-3.0                 &12000               &0.660        &8.1910 &1.50   \\
$\,$ model6           &-8.0                &-3.0                 &11800               &0.665        &8.1939 &1.51   \\
$\,$ model7           &-8.0                &-3.0                 &11700               &0.665        &8.1941 &1.56   \\
$\,$ model8           &-8.0                &-3.0                 &11850               &0.665        &8.1938 &1.72   \\
$\,$ model9           &-8.0                &-3.0                 &12000               &0.655        &8.1825 &1.74   \\
$\,$model10           &-8.0                &-3.0                 &11800               &0.660        &8.1914 &1.75   \\
\hline
ID(R548)              &                    &                     &                    &             &       &       \\
\hline
$\,$ model1           &-8.0                &-3.5                 &12650               &0.645        &8.1714 &2.12   \\
$\,$ model2           &-8.0                &-3.5                 &12600               &0.645        &8.1715 &2.16   \\
$\,$ model3           &-8.0                &-3.5                 &12700               &0.645        &8.1713 &2.19   \\
$\,$ model4           &-8.0                &-3.5                 &12550               &0.645        &8.1716 &2.29   \\
$\,$ model5           &-8.0                &-3.5                 &12750               &0.645        &8.1712 &2.35   \\
$\,$ model6           &-8.0                &-3.5                 &12500               &0.645        &8.1717 &2.50   \\
$\,$ model7           &-8.0                &-3.5                 &12500               &0.650        &8.1799 &2.51   \\
$\,$ model8           &-8.0                &-3.5                 &12550               &0.650        &8.1798 &2.54   \\
$\,$ model9           &-8.0                &-3.5                 &12450               &0.650        &8.1800 &2.57   \\
$\,$model10           &-8.0                &-3.5                 &12800               &0.645        &8.1712 &2.58   \\
\hline
\end{tabular}
\end{center}
\end{table}

\begin{figure}
\begin{center}
\includegraphics[width=9.0cm,angle=0]{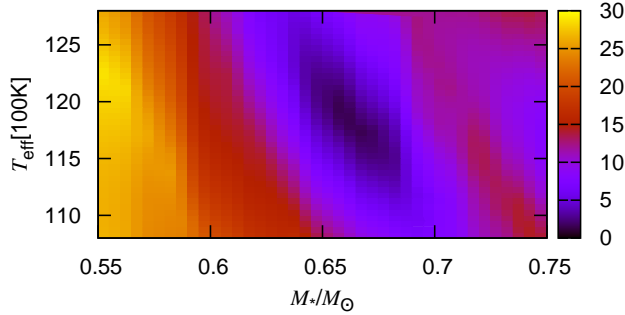}
\end{center}
\caption{Color residual diagram for fitting G117-B15A. The color scale refers to the $\chi$ value.}
\end{figure}

\begin{figure}
\begin{center}
\includegraphics[width=9.0cm,angle=0]{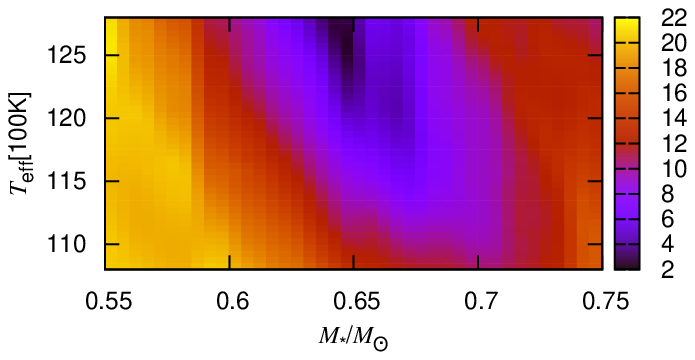}
\end{center}
\caption{Color residual diagram for fitting R548. The color scale refers to the $\chi$ value.}
\end{figure}

The observed modes of G117-B15A and R548 are from Table 5 of Romero et al. (2012). They are listed in Table 2, marked as $P_{\rm obs}$($l$), in this paper. We assume them as $m$ = 0 modes. The modes we calculated are all $m$ = 0 modes. G117-B15A was identified as a DAV star by Richer \& Ulrych (1974). The variability of G117-B15A was confirmed by McGraw \& Robinson (1976) with three modes. Kepler et al. (1982) identified six modes, including the three modes, for G117-B15A. The other three modes can be explained by nonlinear pulsation effects (Brassard et al. 1993). The cancelation effect of neighbouring parts leads to high $l$ modes not easy to be observed. Therefore, the three modes are assumed to be $l$ = 1 modes basically. The assumption is consistent with previous work (Robinson et al. 1995, Bradley 1998, Benvenuto et al. 2002). Bradley (1998) and Benvenuto et al. (2002) fitted them by consecutive $l$ = 1 modes with the 215\,s mode being either a $k$ = 1 or a $k$ = 2 mode. We also assume them to be $l$ = 1 modes. R548 was identified as a harmonically variable white dwarf by Lasker \& Hesser (1970, 1971). Stover et al. (1980) identified the double modes of 213\,s and 274\,s. Kepler et al. (1995) detected other three small-amplitude modes near 187, 320, and 333\,s. We assume the two doublets as $l$ = 1 modes and the three singlets as $l$ = 1 or 2 modes, as shown in Table 2.

Fitting the observed modes in the second column of Table 2, a pre-best-fitting model is selected for each star. The pre-best-fitting model fitting G117-B15A is model1 in Table 3 and the pre-best-fitting model fitting R548 is model7 in Table 3. Then, the grid models are made dense. For G117-B15A, log($M_{H}/M_{*}$) is from -7.0 to -9.0 with steps of 0.5, log($M_{He}/M_{*}$) is from -2.0 to -4.0 with steps of 0.5, $M_{\rm *}$ is from 0.640\,$M_{\odot}$ to 0.680\,$M_{\odot}$ with steps of 0.005\,$M_{\odot}$, and $T_{eff}$ is from 11600\,K to 12200\,K with steps of 50\,K. For R548, log($M_{H}/M_{*}$) is from -7.0 to -9.0 with steps of 0.5, log($M_{He}/M_{*}$) is from -2.0 to -4.0 with steps of 0.5, $M_{\rm *}$ is from 0.630\,$M_{\odot}$ to 0.670\,$M_{\odot}$ with steps of 0.005\,$M_{\odot}$, and $T_{eff}$ is from 12300\,K to 12800\,K with steps of 50\,K. Fitting G117-B15A and R548, the parameters of ten models with smallest $\chi$s are shown in Table 3. The values of $\chi$ are only slightly different. We do not suggest that model1 must have a clear advantage over model2. In mathematical fitting, the value of $\chi$ for model1 is slightly smaller than that for other models. Therefore, model1 is marked as the best-fitting model. Fitting G117-B15A, the ten models with smallest $\chi$s have parameters of log($M_{H}/M_{*}$) = -8.0, log($M_{He}/M_{*}$) = -3.0, $M_{\rm *}$ = 0.655-0.665\,$M_{\odot}$, $T_{eff}$ = 11700-12000\,K, and log$g$ = 8.1825-8.1941. Fitting R548, the ten models with smallest $\chi$s have parameters of log($M_{H}/M_{*}$) = -8.0, log($M_{He}/M_{*}$) = -3.5, $M_{\rm *}$ = 0.645-0.650\,$M_{\odot}$, $T_{eff}$ = 12450-12800\,K, and log$g$ = 8.1712-8.1800. According to the new work of Gianninas, Bergeron \& Ruiz 2011 on the DAV instability strip, the model10 with $T_{eff}$ = 12800\,K and log$g$ = 8.1712 is within the DAV instability strip.

The color residual diagram fitting G117-B15A and R548 is shown in Fig. 11 and 12 respectively. The abscissa is the stellar mass $M_{\rm *}$ and the ordinate is the effective temperature $T_{eff}$. The parameter $\chi$ is represented by color. The values of H and He are the same with the models in Table 3. The calculated periods for the two best-fitting models are shown in Table 2. For G117-B15A, the three observed modes are fitted by $l$ = 1 modes. The radial order $k$ is 1, 2, and 3 respectively. The absolute value of $P_{\rm obs}$ minus $P_{\rm cal}$ is less than 2.00\,s and $\chi$ is 1.32\,s. For R548, the maximal absolute value of $P_{\rm obs}$ minus $P_{\rm cal}$ is 3.14\,s and the parameter $\chi$ is 2.12\,s. In addition, a possible single mode of 217.83\,s with $S/N$ = 3.9 is identified by Giammichele et al. (2015) for R548. The mode of 213\,s is assumed to be an $l$ = 1 mode. The new mode seems to be an $l$ = 2 mode. We assume $m$ = 0 for the mode. The corresponding best-fitting model just has a mode of 218.82\,s ($l$ = 2, $k$ = 4). The absolute value of $P_{\rm obs}$ minus $P_{\rm cal}$ is 0.99\,s for the new mode, which is relatively small. Calculating the six modes for R548 together, the best-fitting model has the value of $\chi$ = 1.97\,s. The six modes are fitted by four $l$ = 1 modes and two $l$ = 2 modes, as shown in Table 2.

  \subsection{Comparing the stellar parameters to previous spectroscopic results and asteroseismological results}

\begin{table}
\caption{The previous spectroscopic results of Bergeron et al. (1995) (B1995 for short), Gianninas, Bergeron \& Ruiz (2011) (G2011 for short), Koester \& Holberg (2001) (K2001 for short), and Giammichele et al. (2015) (G2015 for short).}
\begin{center}
\begin{tabular}{lcccccc}
\hline
References                        &$T_{\rm eff}$        &log$g$         \\
\hline
                                  &[K]                  &               \\
\hline
G117-B15A                         &                     &               \\
\hline
B1995                             &11620$\pm$200        &7.97$\pm$0.05  \\
G2011                             &12600$\pm$193        &8.14$\pm$0.05  \\
K2001                             &12010$\pm$180        &7.94$\pm$0.17  \\
\hline
R548                              &                     &               \\
\hline
B1995                             &11990$\pm$200        &7.97$\pm$0.05  \\
G2011                             &12480$\pm$190        &8.05$\pm$0.05  \\
K2001                             &11865$\pm$170        &7.89$\pm$0.17  \\
G2015                             &12204$\pm$190        &8.012$\pm$0.048\\
\hline
\end{tabular}
\end{center}
\end{table}

In Table 4, we show the previous spectroscopic results. Studying the optical spectra and ultraviolet spectra on DAV stars, Bergeron et al. (1995) suggested that the best value of the mixing length parameter is ML2/$\alpha$=0.6. In our model evolutions by \texttt{WDEC}, the mixing length parameter is also set as ML2/$\alpha$=0.6. Bergeron et al. (1995) reported the spectral result of $T_{\rm eff}$ and log$g$ for 22 DAV stars, including G117-B15A and R548, as shown in Table 4. Based largely on the last published version of McCook \& Sion (1999) catalog, Gianninas, Bergeron \& Ruiz (2011) made a spectroscopic study on over 1100 bright DA white dwarfs (optical spectra) with improved models. The detailed spectral parameters are shown in Table 5 of Gianninas, Bergeron \& Ruiz (2011). Their results for G117-B15A and R548 are selected in Table 4. With ML2/$\alpha$=0.6, Koester \& Holberg (2001) studied 11 DAV stars based on their ultraviolet spectra from IUE or HST. Their results for G117-B15A (WD0921+354) and R548 (WD0133-116) are selected, as shown in Table 4. Spectral work depend strongly on the theoretical treatment of convection. The convective efficiency of ML2/$\alpha$=0.7 is more appropriate for the spectroscopic work of Giammichele et al. (2015) on DAV star GD165 and R548. Tremblay et al. (2013) reported first grid of mean three-dimensional (3D) spectra for DA white dwarfs based on 3D model atmospheres. Taking the 1D/3D corrections of Tremblay et al. (2013) into account, Giammichele et al. (2015) obtained new spectral results for R548.

For G117-B15A, the values of $T_{\rm eff}$ = 11700-12000\,K in Table 3 are consistent with that of B1995, K2001 and slightly smaller than that of G2011 in Table 4. The values of log$g$ = 8.1825-8.1941 in Table 3 are basically consistent with that of G2011 and slightly larger than that of B1995, K2001 in Table 4. For R548, the values of $T_{\rm eff}$ = 12450-12800\,K in Table 3 are consistent with that of G2011 and slightly larger than that of B1995, K2001, G2015 in Table 4. The values of log$g$ = 8.1712-8.1800 in Table 3 are slightly larger than that of B1995, G2011, K2001, and G2015 in Table 4. Overall, the gravitational accelerations are slightly large for our models. The gravitational acceleration is calculated by $\frac{GM}{R^{2}}$. It may be associated with the stellar radius, which will be studied in the future work.

\begin{table}
\caption{The asteroseismological results of Castanheira \& Kepler (2008, 2009) (C2008 and C2009 for short), Bischoff-Kim, Montgomery \& Winget (2008a) (B2008a for short), Romero et al. (2012) (R2012 for short), and Giammichele et al. (2016) (G2016 for short).}
\begin{center}
\begin{tabular}{lcccccccccc}
\hline
References       &log($M_{H}/M_{*}$)  &log($M_{He}/M_{*}$)  &$T_{\rm eff}$       &$M_{\rm *}$     &                  &$\chi$      \\
\hline
                 &                    &                     &[K]                 &[$M_{\odot}$]   &                  &[s]         \\
\hline
G117-B15A        &                    &                     &                    &                &                  &            \\
$l$=1            &                    &                     &                    &                &$k$               &            \\
\hline
C2008            &-7.0                &-2.0                 &12000               &0.615           &1,2,3             &2.85        \\
                 &-5.0                &-2.5                 &11500               &0.750           &2,3,4             &3.84        \\
                 &-7.5                &-3.0                 &12600               &0.710           &1,2,3             &2.94        \\
                 &-8.5                &-3.5                 &11500               &0.850           &1,2,4             &0.74        \\
B2008a           &-6.2                &-2.4                 &11400-12200         &0.650-0.680     &2,3,4             &            \\
                 &(-8.0)-(-7.4)       &-2.4                 &11800-12600         &0.600-0.640     &1,2,3             &            \\
R2012            &(-6.2)-(-5.7)       &-1.6                 &11985$\pm$200       &0.593$\pm$0.007 &2,3,4             &2.14        \\
This paper       &-8.0                &-3.0                 &11900               &0.660           &1,2,3             &1.32        \\
\hline
R548             &                    &                     &                    &                &                  &            \\
$l$=1or2         &                    &                     &                    &                &                  &            \\
\hline
C2009            &-4.5                &-2.0                 &12100               &0.635           &                  &2.64        \\
                 &-5.5                &-2.5                 &11000               &0.790           &                  &3.91        \\
B2008a           &(-7.2)-(-7.7)       &-2.4                 &11700-12600         &0.600-0.650     &                  &            \\
R2012            &-6.0                &-1.6                 &11627$\pm$390       &0.609$\pm$0.012 &                  &3.42        \\
G2016            &-7.45$\pm$0.12      &-2.92$\pm$0.10       &12281$\pm$125       &0.65$\pm$0.02   &                  &0.59        \\
This paper       &-8.0                &-3.5                 &12650               &0.645           &                  &2.12(1.97)  \\
\hline
\end{tabular}
\end{center}
\end{table}

In Table 5, we show the asteroseismological results of Castanheira \& Kepler (2008, 2009), Bischoff-Kim, Montgomery \& Winget (2008a), Romero et al. (2012), and Giammichele et al. (2016). Castanheira \& Kepler (2008, 2009) evolved grids of DAV star models by \texttt{WDEC} with homogeneous C/O core compositions and did asteroseismological study on over 80 DAV stars, including G117-B15A and R548. They obtained four 'best-fitting' models fitting G117-B15A, and two 'best-fitting' models fitting R548. Bischoff-Kim, Montgomery \& Winget (2008a) evolved grids of DAV star models by \texttt{WDEC} based on chemical composition profiles of a fiducial model and models with sharper C/He transition zone, as shown in Fig. 2 of Bischoff-Kim, Montgomery \& Winget (2008a). They did asteroseismological study on G117-B15A and R548. Their best-fitting models with 'thick' H, thin H for G117-B15A and thin H for R548 are displayed in Table 5. Romero et al. (2012) made fully evolutionary DA white dwarf models characterized by detailed chemical profiles by \texttt{LPCODE} evolutionary code. Based on those DAV star models, they did asteroseismological study on 44 bright DAV stars, including G117-B15A and R548. In addition, Giammichele et al. (2016) did asteroseismological study on GD165 and R548 based on parameterized static models. Their best fitting model is shown in Table 5. According to those best-fitting models in Table 5, the values of corresponding $\chi$ are calculated. All the values are smaller than 4.00\,s. We notice that the best-fitting models depend on the stellar evolution code, such as the input core composition profiles.

For G117-B15A, the three modes are fitted by the best fitting model (model1 in Table 3) with $k$ = 1, 2, 3 mode ($l$ = 1) respectively. They are fitted by the model of Romero et al. (2012) with $k$ = 2, 3, 4 mode ($l$ = 1) respectively. That is why we obtain a thin H mass model and they have a relatively thick H mass model. The discussion about thin H mass or thick H mass had been reported in detail by Bradley (1998) and Benvenuto et al. (2002). Bischoff-Kim, Montgomery \& Winget (2008a) obtained a group of relatively 'thick' H mass models ($k$ = 1, 2, 3) and a group of thin H mass models ($k$ = 2, 3, 4). The three observed modes are fitted by Castanheira \& Kepler (2008) with $k$ = 1, 2, 3 modes for thin H models, except a large mass ($M_{\rm *}$ = 0.850\,$M_{\odot}$) one, and with $k$ = 2, 3, 4 modes ($l$ = 1) for thick H model. For the model of $M_{\rm *}$ = 0.850\,$M_{\odot}$, the three modes are fitted by $k$ = 1, 2, 4 modes, as shown in the sixth column in Table 5.

For R548, the five modes are fitted by four $l$ = 1 modes and one $l$ = 2 mode for the best fitting model (model1 in Table 3), as shown in Table 2. The mode of 187\,s is fitted by an $l$ = 2 and $k$ = 3 mode. The other modes are fitted by $l$ = 1 and $k$ = 1, 2, 3, 4 mode respectively. The best fitting model fitting R548 is similar to the best fitting model fitting G117-B15A. The two best-fitting models are thin H models with log($M_{H}/M_{*}$) = -8.0, as shown in Table 3. Castanheira \& Kepler (2009) and Romero et al. (2012) obtained relatively thick H mass models fitting R548. For the first model of Castanheira \& Kepler (2009), the modes of 187\,s and 333\,s are fitted by $l$ = 2 and $k$ = 3, 9 modes. The other three modes are fitted by $l$ = 1 and $k$ = 1, 2, 3 modes. For the second model of Castanheira \& Kepler (2009), the mode of 187\,s is fitted by an $l$ = 2 and $k$ = 3 mode while the other four modes are fitted by $l$ = 1 and $k$ = 2, 3, 4, 5 modes. For the best-fitting model of Romero et al. (2012), the modes of 318\,s and 333\,s are fitted by $l$ = 2, $k$ = 8, 9 modes. The other three modes are fitted by $l$ = 1 and $k$ = 1, 2, 3 modes. For the models of Bischoff-Kim, Montgomery \& Winget (2008a), the modes of 187\,s and 334\,s are fitted by $l$ = 2 and $k$ = 4, 8 modes. The other three modes are fitted by $l$ = 1 and $k$ = 1, 2, 4 modes. The mode identifications of our best-fitting model are not consistent with their results.

Giammichele et al. (2016) did asteroseismological study on R548 based on parameterized static models from the optimization package LUCY (Charpinet et al. 2015). They obtained a best-fitting model with thin H mass. Their best-fitting model parameters are log($M_{\rm H}/M_{\rm *}$) = -7.45$\pm$0.12, log($M_{\rm He}/M_{\rm *}$) = -2.92$\pm$0.10, $M_{\rm *}$ = 0.65$\pm$0.02\,$M_{\odot}$, $T_{\rm eff}$ = 12281$\pm$125\,K, and log$g$ = 8.108$\pm$0.025. Their asteroseismological study on R548 is basically consistent with their spectroscopic work with $T_{\rm eff}$ = 12204$\pm$190\,K, and log$g$ = 8.012$\pm$0.048. For their best-fitting model, the modes of 187\,s and 217\,s are fitted by $l$ = 2 and $k$ = 3, 4 modes. The other four modes are fitted by $l$ = 1 and $k$ = 1, 2, 3, 4 modes. The mode identifications of our best-fitting model are the same with that of the best-fitting model of Giammichele et al. (2016). In addition, the values of H, He, $T_{\rm eff}$, $M_{\rm *}$ for our best-fitting model are consistent with or close to that of the best fitting model of Giammichele et al. (2016), as shown in Table 5.

  \subsection{Comparing the the rates of period change to the observed values obtained through O-C method}

\begin{table}
\caption{Comparing the calculated values of $\dot{P}$ of the two best-fitting models in Table 3 to the observed values of that obtained through O-C method for G117-B15A and R548.}
\begin{center}
\begin{tabular}{lllll}
\hline
Star         &$P_{\rm obs}$ & $\dot{P}{\rm obs}$ & $P_{\rm cal}$($l$,$k$)   &$\dot{P}{\rm cal}$ \\
\hline
             &[s]         & [$10^{-15}$s/s]      & [s]                      &[$10^{-15}$s/s]    \\
\hline
G117-B15A    &215.20        & 4.19$\pm$0.73      &213.86(1,1)               &4.19               \\
             &270.46        &                    &270.92(1,2)               &2.40               \\
             &304.05        &                    &305.85(1,3)               &3.96               \\
\hline
R548         &187.28        &                    &190.42(2,3)               &2.25               \\
             &212.95        & 3.3$\pm$1.1        &214.41(1,1)               &4.21               \\
             &217.83        &                    &218.82(2,4)               &3.75               \\
             &274.51        &                    &272.48(1,2)               &2.15               \\
             &318.07        &                    &317.97(1,3)               &5.39               \\
             &333.64        &                    &331.13(1,4)               &4.02               \\
\hline
\end{tabular}
\end{center}
\end{table}

Based on the observations from 1974 to 2010 on G117-B15A, Kepler (2012) reported the rate of period change for 215\,s as (4.89 $\pm$ 0.53) $\times$ 10$^{-15}$\,s/s. The proper motion correction is (-0.7 $\pm$ 0.2) $\times$ 10$^{-15}$\,s/s. Namely, the value of $\dot{P}{\rm obs}$ is (4.19 $\pm$ 0.73) $\times$ 10$^{-15}$\,s/s for the 215\,s mode of G117-B15A, as shown in Table 6. For our best-fitting model, the mode of 213.86\,s ($l$=1,$k$=1) is used to fit the 215\,s mode. We calculate the rate of period change for the mode of $l$ = 1 and $k$ = 1. The value of $\dot{P}{\rm cal}$ is just 4.19 $\times$ 10$^{-15}$\,s/s, which is exactly consistent with the observed value. For the mode of 213.86\,s ($l$=1,$k$=1), there are 96.44\% of kinetic energy distributed in He layer.

The DAV star R548 has been observed from 1970 November to 2012 January. Using 41 years of time-series photometry, Mukadam et al. (2013) calculated the rate of period change for the 213\,s mode. Taking the correction of proper motion into account, Mukadam et al. (2013) obtained $\dot{P}{\rm obs}$ = (3.3 $\pm$ 1.1) $\times$ 10$^{-15}$\,s/s for the 213\,s mode. For our best-fitting model, the mode of 214.41\,s ($l$=1,$k$=1) is used to fit the 213\,s mode of R548. We calculate the rate of period change for the mode of $l$ = 1 and $k$ = 1. The value of $\dot{P}{\rm cal}$ is 4.21 $\times$ 10$^{-15}$\,s/s, which is consistent with the observed value. Fitting the mode of 213\,s, there are 95.61\% of kinetic energy distributed in the He layer for the mode of 214.41\,s ($l$=1,$k$=1). Giammichele et al. (2016) obtained the calculated value of (2.87-2.91) $\times$ 10$^{-15}$\,s/s for the corresponding mode. It is consistent with the observed value. If the value of $\dot{P}{\rm cal}$ is smaller than the value of $\dot{P}{\rm obs}$, the results can be used to constrain the mass of axion (Bischoff-Kim, Montgomery \& Winget 2008b, C$\acute{o}$rsico et al. 2012a, C$\acute{o}$rsico et al. 2012b, Mukadam et al. 2013, C$\acute{o}$rsico et al. 2016). However, the two modes of our best-fitting models fitting the 215\,s mode of G117-B15A and the 213\,s mode of R548 are trapped or partly trapped in the He layer. They have relatively large rates of period change. The values are consistent with the observed values.

\section{Discussion and conclusions}

In this paper, we introduce a method of evolving DAV stars. Groups of $MS$ stars are evolved to be $WD$ stars by a module named 'make\_co\_wd' from \texttt{MESA} (version 6208). The core compositions are took out and added into \texttt{WDEC}. With historically viable core compositions, grids of DAV star models are evolved by \texttt{WDEC}, taking the element diffusion into account (Thoul, Bahcall \& Loeb 1994, Su et al. 2014). Those DAV star models are used to study the rate of period change in DAV stars. Then, we do asteroseismological study on DAV stars G117-B15A and R548 based on the grids of DAV star models. At last, we compare the calculated rates of period change with the observed values through O-C method.

Studying the rate of period change in DAV stars, we try to discuss the effect of different values of H atmosphere mass, He layer mass, stellar mass $M_{\rm *}$, and effective temperature $T_{\rm eff}$. Different thickness of H atmosphere and He layer will affect the rate of period change by mode trapping effect. Modes trapped or partly trapped in C/O core or He layer have relatively large rate of period change. But modes trapped or partly trapped in H atmosphere have relatively small rate of period change. This is due to the fast cooling process of the C/O core and He layer. The cooling process dominates the rate of period change in DAV stars. The results are consistent with previous work (Bradley \& Winget 1991, Bradley, Winget \& Wood 1992). The rate of period change is sensitive to the stellar mass and the effective temperature. A large $M_{\rm *}$ DAV star has a long process of cooling down and then a small rate of period change (Bradley \& Winget 1991, Bradley 1996). A high $T_{\rm eff}$ DAV star has a fast time of cooling down and then a large rate of period change. The effect of different values of $M_{\rm *}$ and $T_{\rm eff}$ is obvious for the long-period modes.

Based on the observed modes of G117-B15A and R548 from Romero et al. (2012), we evolve grids of DAV star models and do asteroseismological study on the two DAV stars. A best-fitting model is selected for each star by selecting a minimal value of $\chi$. The two DAV stars have short-period modes observed, as shown in Table 2, and therefore they should have hot effective temperatures (Clemens 1993, Mukadam et al. 2006). The corresponding two best-fitting models are really hot DAV star models, with $T_{\rm eff}$ = 11900\,K for G117-B1A and $T_{\rm eff}$ = 12650\,K for R548. For the fitting results, the maximal absolute value of observed mode minus calculated mode is 3.14\,s. The value of $\chi$ is respectively 1.32\,s fitting G117-B15A and 2.12\,s fitting R548. In addition, the new observed mode of 217.83\,s (Giammichele et al. 2015) for R548 can also be fitted by a mode of 218.82\,s ($l$=2, $k$=4) for the best-fitting model.

The DAV stars G117-B15A and R548 are observationally similar. The best-fitting model fitting R548 is similar to the best-fitting model fitting G117-B15A. The calculated rate of period change for the mode of 213\,s of R548 is also similar to that for the mode of 215\,s of G117-B15A. Based on the two best-fitting models, the mode identifications ($l$, $k$) of the observed modes for G117-B15A and R548 are consistent with previous work for G117-B15A (thin H model of Bradley 1998, Benvenuto et al. 2002, Castanheira \& Kepler 2008, Bischoff-Kim, Montgomery \& Winget 2008a) and R548 (Giammichele et al. 2016). In addition, fitting G117-B15A, the calculated rate of period change is exactly consistent with the corresponding observed one through O-C. Fitting R548, the calculated rate of period change is consistent with the corresponding observed one. Basically, both the observed modes and observed rates of period change obtained through the O-C method can be fitted. The results greatly increase our confidence on the asteroseismological study on DAV stars. The results indicate that the method of evolving DAV stars (MESA(core)+WDEC(diffusion)) is feasible.

\section{Acknowledgment}

The work is supported by the NSFC of China (Grant Nos. 11563001, 11663001), the Yunnan Applied Basic Research Project (2015FD044, 2015FD045), the Open Research Program of key Laboratory for the Structure and Evolution of Celestial Objects, Chinese Academy of Sciences (OP201502, OP201507), and the Research Fund of Chuxiong Normal University (XJGG1501, 14XJGG03). We are very grateful to Y. Li, T. Wu, X. H. Chen, and J. Su for their kindly discussion and suggestions.

\label{lastpage}
\end{document}